\documentclass[12pt]{iopart}

\usepackage{iopams} 
\usepackage{graphicx} 
\begin{document}

\title[Squeezed light at the Rubidium D1 line]{Squeezed light for bandwidth limited atom optics experiments at the Rubidium D1 line}

\author{G. H\'etet, O.~Gl\"ockl, K. A. Pilypas, C.C. Harb, B.C. Buchler, H.-A. Bachor, P.K. Lam}

\address{ARC COE for Quantum-Atom Optics, Australian National University, Canberra, ACT 0200, Australia}
\ead{ping.lam@anu.edu.au}

\begin{abstract}
We report on the generation of more than 5 dB of vacuum squeezed light at the Rubidium D1 line (795 nm) using periodically poled KTiOPO$_{4}$ (PPKTP) in an optical parametric oscillator. We demonstrate squeezing at low sideband frequencies, making this source of non-classical light compatible with bandwidth limited atom optics experiments. When PPKTP is operated as a parametric amplifier, we show a noise reduction of 4 dB stably locked within the 150 kHz-500 kHz frequency range. This matches the bandwidth of Electromagnetically Induced Transparency (EIT) in Rubidium hot vapour cells under the condition of large information delay. 
\end{abstract}

\maketitle

The control of quantum states of light is of great interest for quantum communication purposes. Many quantum information protocols rely on the possibility of coherently delaying and storing the information carried by a laser beam. This can for example be achieved using Electromagnetically Induced Transparency (EIT). Experimental and theoretical studies have shown that information can be delayed within a narrow bandwidth when light interacts with atoms in a lambda configuration \cite{hau,akamatsu,peng,hsu}.  One step further is the delay of quantum information, carried by squeezed or entangled light states, through such a system. To achieve this goal a source of light with non-classical photon statistics at low sideband frequencies tuned to atomic transitions is required. Other applications of low frequency squeezing in atom optics include the generation of continuous variable entanglement between an atom laser beam and an optical field. This can be realized via outcoupling of atoms from a Bose-Einstein condensate using squeezed light in a Raman transition \cite{haine}. 

Many of these atom optics experiments are performed using transitions at the Rubidium D1 line at 795 nm. The generation of squeezing at these wavelengths has therefore been the subject of many experimental efforts involving either atomic interactions in Rubidium or by using the $\chi^{(2)}$ interaction in non-linear media. Squeezing via self-rotation in thermal vapour cells \cite{ries,hsu2} was shown to be a challenge,  since atomic noise reduces the amount of squeezing that could be observed. Relative intensity squeezing of up to 3.5 dB generated by Four-wave mixing in Rb vapour has been shown recently \cite{McCormick}. When using $\chi^{(2)}$ non-linear media for the squeezed light generation, it is difficult to identify materials that have high non-linearities and negligible passive losses, as the second harmonic of the Rb D1 line, 397.5 nm light, approaches the UV cut-off wavelength of many non-linear optical media. Periodically poling on the other hand allows to tailor the properties of non-linear media so that quasi phase matching can be achieved for any desired optical frequency. Early experiments used periodically poled material in waveguide form which offers a high non-linearity and an extended interaction length. However, losses in waveguides are higher than in bulk material due to technological issues. To date, 0.9 dB of squeezing at 795 nm have been reported using waveguides \cite{akamatsu}. KTiOPO$_{4}$ (KTP) presents a high non-linearity together with a good transmission at 397.5 nm which makes periodically poled KTP (PPKTP) a good candidate for experiments at the D1 Rubidium line. More recent experiments showed the great potential of squeezed state generation in parametric downconversion using PPKTP. More than 7 dB of quadrature squeezing has been reported at 860 nm \cite{Suzuki}, while at 795 nm, to date 2.75 dB of squeezed vaccum has been observed \cite{tanimura}. 

Here, we report on improvements on the generation of squeezing at Rubidium wavelengths and show more than 5 dB of quantum noise suppression of a vacuum field using optical parametric oscillation (OPO). By operating our system with a seed beam of low power (i. e. by running the OPO as an amplifier, OPA), we locked the system to deamplification and show more than 4 dB of amplitude quadrature squeezing down to 150 kHz which makes this non-classical light source suitable for atom optics experiments. 
 
\begin{figure}[htbp]
\begin{center}
\includegraphics[width=10cm]{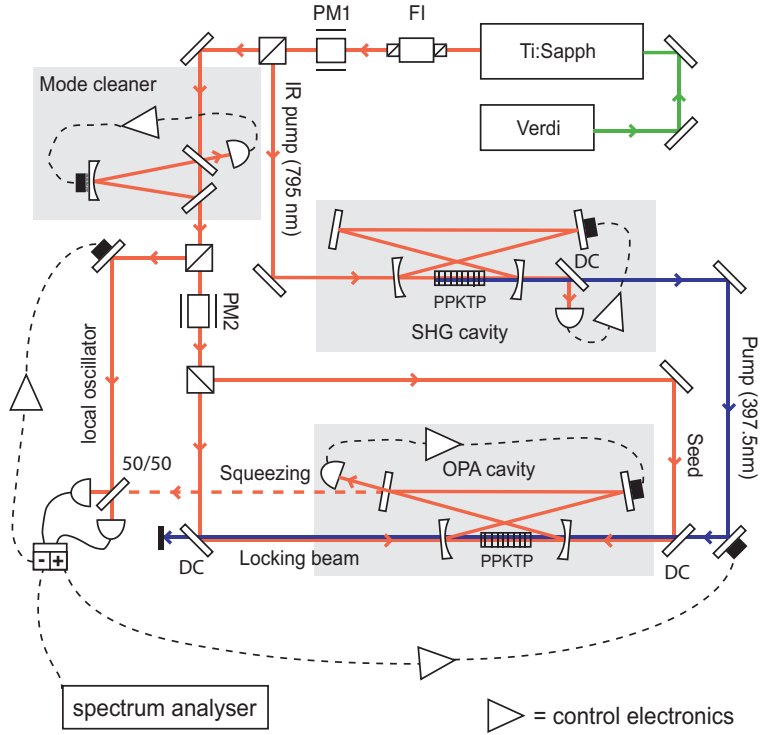}
\caption{Schematic of the experimental setup. FI: Faraday isolator, PM:
electro optic phase modulator, DC: Dichroic mirror, OPA: sub-threshold
optical parametric amplifier. The control electronics typically consists of a mixer, a low pass filter, a proportional Integral differential (PID) controller, and high voltage amplifier.}
\label{setup}
\end{center}
\end{figure}

In our experiment (set-up see Fig.(\ref{setup})) we used a Ti:Sapphire laser (Coherent MBR)  tuned to 795 nm to pump a second-harmonic generator (SHG), to seed the OPA, and to provide a local oscillator for the homodyne detection system. For both the SHG and the OPA bow-tie cavities with PPKTP crystals of length 20 mm and cross-section 1$\times$2 mm$^2$ were used. The cavities had a 600 mm round trip perimeter and two curved mirrors (100 mm radius of curvature) giving a waist of 40 $\mu$m inside the crystal. With these parameters a single pass efficiency of $\Gamma=\mathrm{P}_{2\omega}/\mathrm{P}_\omega^2\sim$2.7$\times10^{-2}$ W$^{-1}$ for the PPKTP crystal was measured. Here, $\mathrm{P}_\omega$ and $\mathrm{P}_{2\omega}$ refer to the pump power at the fundamental and the second harmonic field respectively. The waist size was chosen to minimize thermal effects whilst still providing enough UV light being generated in the SHG. To simplify mode-matching, the OPA cavity was of identical geometry to the SHG. By varying the temperature of the crystals between 20 and 50$^\circ$C we were able to produce UV light over a range of 1.5 nm around 795.3 nm. The SHG cavity was pumped with 300 mW of infrared light via a 82 \% reflectivity flat input coupling mirror. At a phase matching temperature of around 20$^\circ$C, UV light at 397.5 nm was efficiently generated. The maximum second harmonic conversion efficiency reached in that regime was up to 50$\%$. The locking was done by detecting the transmitted pump using Pound Drever Hall (PDH) techniques \cite{PDH} by applying a phase modulation on the pump at 10.4 MHz using PM1. The second harmonic light was coupled out from the SHG through one of the curved mirrors and was mode matched to the OPA cavity. We observed that our PPKTP crystal is prone to grey tracking when it is used for SHG at higher power density levels \cite{graytrack}. We noticed a degradation in the efficiency of the frequency doubler as well as a distortion of the mode shape of the second harmonic output field after operating the SHG over a longer period of time. The effect of grey tracking was partly reversed by slowly heating the crystal up to around 120$^\circ$C and baking it for a period of several days as suggested in Ref. \cite{graytrack_recovery}. To minimize grey tracking effects, we restricted the amount of UV light produced to around 50 mW. Operation in this regime also minimized photo thermal effects in the SHG cavity and allowed for a more stable locking. While grey tracking was observed in SHG with its relative high power density levels involved, we observed no deterioration of the non-linear crystal when it was used in a sub-threshold OPA.  

A mode cleaning cavity was used to generate a TEM$_{00}$ beam. This facilitated the mode matching of the beams into the OPA cavity and ensured a high interference contrast in the homodyne detection. As the OPA cavity was highly impedance mismatched, deriving an error signal from the reflected seed was more difficult and yielded non optimum locking. The OPA cavity length was therefore locked on resonance using an auxiliary beam propagating in the opposite direction to the pump beam as shown in Fig. \ref{setup}. The modulator PM2 provided phase modulation sidebands at $5.6$ MHz allowing the generation of a PDH error signal for the OPA cavity. Furthermore, locking the cavity via the counter propagating beam allowed for the stable generation of vacuum squeezing as the cavity length can be locked independently from the seed.

The OPA was pumped through one of the curved mirrors. After optimizing the mode matching of the pump into the cavity we measured a threshold of around 25 mW, with an output coupler of 95\%. The total losses inside the OPA are then estimated to be around 0.5\%. For the actual squeezed state generation, we changed the reflectivity of the output coupling mirror to 92\%. This enhances the squeezing escape efficiency, which we calculate to be 93\%. The corresponding theoretical threshold is now 68 mW.  We pumped our OPA with 40 mW. A parametric gain of around 10 was observed in that regime. In the first step we blocked the seed, thus running the OPA as an OPO to generate a squeezed vacuum state. The squeezing was measured with a homodyne detection scheme. The mode matching between the local oscillator and the OPA output was 97$\%$ and the photodiode quantum efficiency was around 95$\%$. The overall efficiency, also taking into account the escape efficiency of the squeezing from the cavity, yielded 83$\%$.

\begin{figure}[htbp]
\begin{center}
\includegraphics[width=8.5cm]{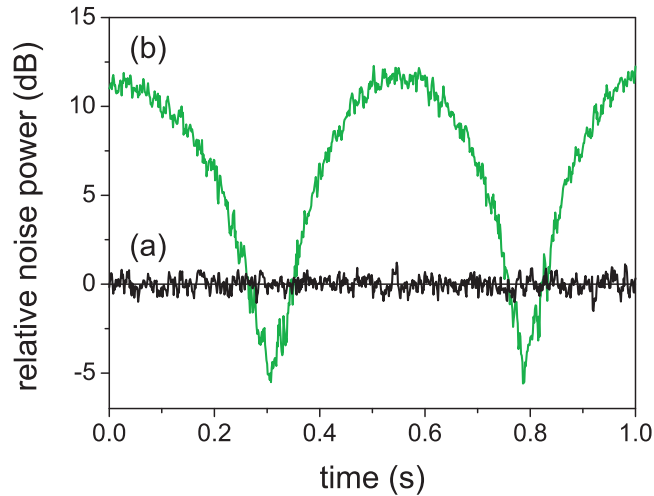}
\caption{Quantum noise levels for squeezed vacuum generation. (a) Quantum noise level and (b) squeezing trace when scanning the local oscillator phase. Noise levels
are displayed as the relative power compared to the shot noise level. The settings on the spectrum analyzer were, zero-span mode at 400 kHz, resolution bandwidth
= 30 kHz, video bandwidth = 100 Hz.}
\label{macdonalds}
\end{center}
\end{figure}

Figure \ref{macdonalds} shows the homodyne detection signal measured in zero span mode at 400 kHz with a spectrum analyser, when the phase of the local oscillator was scanned. The resolution bandwidth was 30 kHz and the video bandwidth 100 Hz. This curve shows a noise reduction of 5.2 dB $\pm$ 0.4 dB below the quantum noise limit when correcting for electronic noise (which is 10 dB below the quantum noise level defined by our local oscillator beam), the anti-squeezing level is 12 dB $\pm$ 0.4 dB. 

\begin{figure}[htbp]
\begin{center}
\includegraphics[width=8.5cm]{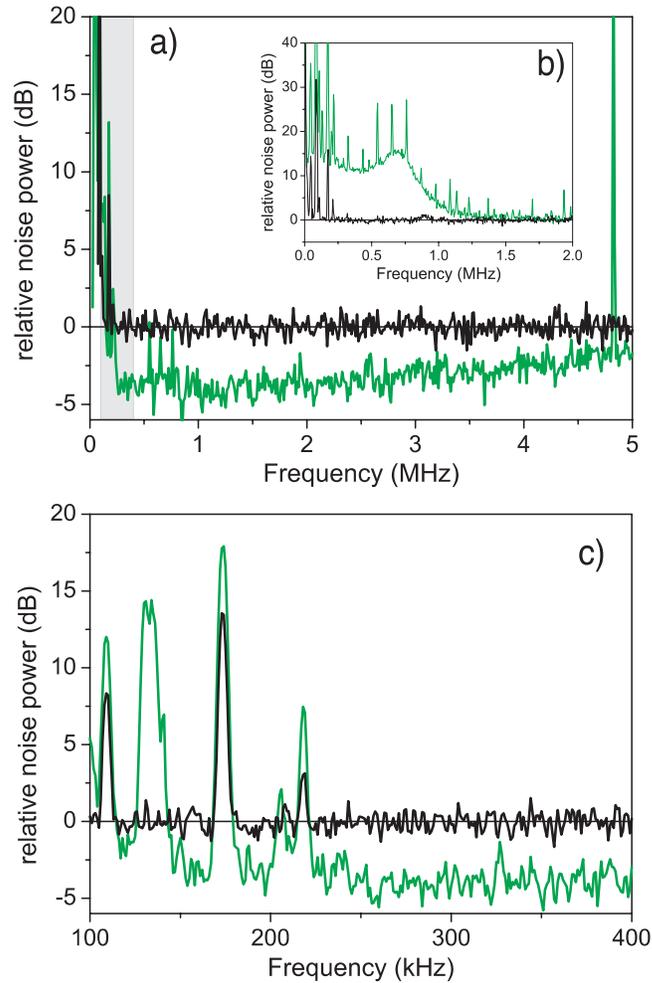}
\caption{Squeezing spectra observed from the OPA, normalised to the quantum noise limit. a) broad frequency range, b) laser noise at 1 mW, c) squeezing in the low frequency range 100 kHz-400 kHz (shaded area in a). The resolution bandwidth was 3 kHz for these measurements, the video bandwidth was 30 Hz (300 Hz) for the low (broad) frequency range.}
\label{locked}
\end{center}
\end{figure}

Next, we seeded the OPO to run it as a parametric amplifier. To generate squeezing at a particular quadrature, the phases of the pump beam and of the local oscillator with respect to the seed beam need to be controlled. We chose to lock the pump phase to deamplify the seed and thus generate an amplitude squeezed beam. The locking signal was derived from the phase modulation signal of the seed beam at 5.6 MHz . This modulation signal was transmitted through the cavity via the seed beam and measured on the two homodyne detectors. The sum signal of the detectors was then demodulated and low pass filtered thus providing an error signal to lock the pump phase with respect to the seed beam. As we wanted to measure the amplitude quadrature of the squeezed beam, the homodyne detector needed to be locked to the bright/dark fringe. The same phase modulation signal at 5.6 MHz which was used for locking to deamplification was also used to lock the homodyne detection. The locking signal was obtained via the difference signal of the photocurrents of the two detectors. Figure \ref{locked}.a) shows the evolution of the squeezing spectrum measured at sideband frequencies within the cavity bandwidth. The quantum noise suppression gets more efficient at low frequencies as expected. At sideband frequencies below 1 MHz, technical noise is coupled into the system.

The ultimate sources of noise limiting the low frequency performance of the squeezer are the noise on the seed and, noise of the pump coupled onto the squeezed beam when the OPA is seeded. At such low frequencies, most of the lasers show technical noise (see inset b) in Fig. \ref{locked}). Different techniques have been proposed and implemented to overcome these issues. One approach relies on interferometric cancellation of common mode noise, either by interference of two squeezed beams from two OPAs which were seeded by the same laser on a symmetric beam splitter, \cite{bowen} or by placing a squeezer inside a Mach-Zehnder configuration \cite{schnabel}. Alternatively the seed power level can be reduced. This minimises the technical noise at low frequencies on the seed beam, but also reduces the coupling of noise from the pump onto the squeezed field \cite{mckenzie}. In the limit of zero power in the seed, no noise is coupled to the squeezed field, however, to lock the local oscillator to a particular quadrature, a different locking technique such as quantum noise locking \cite{mckenzie_noiselock} has to be employed. Quantum noise locking requires the detection of the squeezing spectrum over a large bandwidth, which is incompatible with bandwidth limited atomic systems. To lock the squeezer to deamplification, we therefore reduced the power level of the seed beam to minimise the noise coupling onto the squeezed beam while still being able to generate large enough error signals. As the locking stability of the system, on the other hand, relies on the total power in the seed, a compromise between seed power and coupling of low frequency noise into the squeezed beam and locking stability needed to be found. In Fig. \ref{locked}.c), the low frequency part of the squeezing spectrum is plotted. More than 3 dB of amplitude squeezing were measured at sideband frequencies down to 150 kHz. The results presented here were obtained with a seed power of 2 $\mu$W whilst we were still able to lock stably the set-up for a few minutes.

Our main motivation for the generation of squeezing at atomic wavelengths is to study the performance of EIT in Rubidium to delay and store quantum information. 
The amount of delay possible in such atomic systems at a given sideband frequency is governed by the dispersion properties, which in turn are linked via the Kramers Kronig relation to the transmission bandwidth.
For a quantum delay line to be efficient, passive losses must be avoided. Thus, useful sideband frequencies where the information is encoded are limited to the sub-megahertz regime. With our squeezing source at hand, it should be possible to demonstrate the delay of quantum information at a sideband frequency of 150 kHz by $20 \mu$s with only moderate losses of 20\% \cite{peng}. However, recent sudies showed that additional sources of decoherence might limit the performance of such systems further, i. e. restrict the transmission and also add extra noise \cite{hsu}. Our aim was therefore to produce large stably locked squeezing at such low (sub-MHz) frequencies to probe the capabilities of EIT as a quantum information delay line.  

In conclusion, we have demonstrated $5.2\pm0.4$ dB of vacuum squeezing at the Rubidium D1 line using optical parametric oscillation. PPKTP was used as the non-linear crystal for frequency doubling and down conversion. We were able to stably lock the OPA to generate amplitude squeezing in the frequency range compatible with bandwidth limited atom optics experiments. Around 4 dB of quantum noise suppression was achieved in a frequency range from 150 kHz to 500 kHz. The low frequency performance is currently limited by laser noise which couples onto the squeezed beam. The use of an intensity stabilized laser source and the application of an external intensity noise eater should lead to further improvements of the low frequency squeezing. Already now, the squeezing generated gives us the possibility to efficiently examine the quantum performance of EIT-based delay lines.   

We would like to thank M. T. L. Hsu, N. B. Grosse and K. McKenzie for useful discussions. We acknowledge funding from the Defense Science and Technology Organisation and the involvement of D. Pulford.
This work was funded by the ARC Centre of Excellence for Quantum-Atom Optics.

\end{document}